\documentclass[conference]{IEEEtran}

\renewcommand{\paragraph}[1]{\vspace{0.1cm}\noindent{\bf #1.}}
\newcommand{\myparagraph}[1]{\vspace{0.1cm}\noindent{\it #1.}}
\usepackage{graphicx}
\usepackage{xcolor}
\usepackage{booktabs}
\usepackage{float}
\usepackage{tikz}
\usepackage{xspace}
\usepackage{paralist}
\usepackage[hyphens]{url}
\usepackage{multirow}
\usepackage[font={small},skip=3pt,belowskip=-7pt]{caption}
\usepackage[bookmarks=true,draft]{hyperref}

\usepackage{array}
\usepackage{rotating}
\usepackage{makecell}
\usepackage{colortbl}

\usepackage{amsmath,amssymb,amsfonts}

\renewcommand{\tablename}{Tab.}
\renewcommand{\figurename}{Fig.}

\newcommand{\specialcell}[2][c]{\begin{tabular}[#1]{@{}l@{}}#2\end{tabular}}

\begin{document}

	\title{SoK: Cryptocurrency Wallets -- A Security Review and Classification based on Authentication Factors}

	\author{\IEEEauthorblockN{Ivan Homoliak}
	\IEEEauthorblockA{Brno University of Technology\\
		Faculty of Information Technology\\
		ihomoliak@fit.vutbr.cz}
	\and
	\IEEEauthorblockN{Martin Pere\v{s}\'{i}ni}
	\IEEEauthorblockA{Brno University of Technology\\
		Faculty of Information Technology\\
		iperesini@fit.vutbr.cz}
	}

\maketitle

\begin{abstract}
		In this work, we review existing cryptocurrency wallet solutions with regard to authentication methods and factors from the user's point of view.
		In particular, we distinguish between authentication factors that are verified against the blockchain and the ones verified locally (or against a centralized party).
		With this in mind, we define notions for $k-factor$ authentication against the blockchain and $k-factor$ authentication against the authentication factors.
		Based on these notions, we propose a classification of authentication schemes.
		We extend our classification to accommodate the threshold signatures and signing transactions by centralized parties (such as exchanges or co-signing services).
		Finally, we apply our classification to existing wallet solutions, which we compare based on various security and key-management features.
	\end{abstract}

	\section{Introduction}\label{sec:intro}
Cryptocurrencies promise to revolutionize many fields and businesses, and indeed, they have been successful beyond expectations.
Decentralized cryptocurrency platforms allow users to conduct monetary transfers, write smart contracts, make loans, and participate in predictive markets while benefiting from features that centralized platforms could not guarantee, such as decentralization, censorship resistance, integrity, transparency, 100\% availability, etc.
Cryptocurrencies utilize native crypto-tokens (a.k.a., coins), which can be transferred in transactions authenticated by private keys that belong to crypto-token owners.
The users owning crypto-tokens interact with the cryptocurrency through a wallet software that manages their private keys.

Unfortunately, there are many cases of stolen keys and attacks on e.g., brain wallets~\cite{2016-brainwallets,courtois2016speed}, cryptocurrency exchanges~\cite{binance-hack-2019,bitmart-hack-2021,ronin-hack-2022,coinex-hack-2023}, and even hardware wallets~\cite{kraken-trezor-2019} with Cortex M3/M4 micro-controllers (such as Trezor, Ledger, KeepKey, etc.).
Such cases have brought the research community's attention to the security issues related to key management in cryptocurrencies~\cite{eskandari2018first,goldfeder2015securing,2015-Bitcoin-SOK,erinle2023sok}.

\subsection{Existing Surveys}\label{sec:surveys}
The works of Eskandri et al.~\cite{eskandari2018first} and Bonneau et al.~\cite{2015-Bitcoin-SOK}  proposed a categorization of cryptocurrency wallets with regard to key management.
The works of Huoy et al.~\cite{2023-Huoy-Sok-wallets-security} and Erinle et al.~\cite{erinle2023sok} reviewed broad vulnerabilities and defenses related to wallets while they specified their categorizations.
Suratkar et al.~\cite{suratkar2020cryptocurrency} focused on a review of wallets with regard to supported coins, anonymity, cost, platform support, key management, and recovery methods. Karantias~\cite{2020-Karantias-sok-wallets} reviewed a few categories and instances of wallets with regard to the verification of transactions,
privacy, communication complexity, and censorship.

\subsection{Authentication Schemes of Wallets \& Security Issues}\label{sec:wallets-review}
We base on the works of Eskandri et al.~\cite{eskandari2018first} and Bonneau et al.~\cite{2015-Bitcoin-SOK}, who focus on the key management approaches, and we review and extend their categorizations.
Private keys are encrypted with selected passwords in \emph{pass\-word-protected wallets}.
Unfortunately, users often choose weak passwords that can be brute-forced if stolen by malware \cite{2015-CCSM-SecureWorks}; optionally, such malware may use a keylogger \cite{2015-Bitcoin-SOK,2017-keylogger-bc-malware}.
Another similar option is to use \emph{password-derived wallets} that generate keys based on the provided password.
However, they also suffer from the possibility of weak passwords \cite{courtois2016speed}.
\emph{Hardware wallets} enable only the signing of transactions, without revealing the private keys stored on the device.
However, these wallets do not provide protection from an attacker with full access to the device \cite{kraken-trezor-hack,kraken-keepkey-hack,donjon-ellipal-hack}, and more importantly, some wallets do not have a secure channel for informing the user about the details of a transaction being signed (e.g., \cite{ledger-nano}) may be exploited by malware targeting IPC mechanisms~\cite{bui2018man}.

A popular option for storing private keys is to deposit them into  \emph{server-side hosted} (i.e., custodial) \emph{wallets} and currency-exchange services \cite{CoinbaseWallet,binance-exchange,poloniex-exchange,kraken-exchange,luno-wallet,paxful-wallet}.
In contrast to the previous categories, server-side wallets imply trust in a provider (storing the private keys of the wallets), which is a potential risk of this category.
Due to many cases of compromising server-side wallets \cite{2018-coindesk-bithumb,2014-Mt-Gox,2016-Bitfinex-hack,moore2013beware,binance-hack-2019} or fraudulent currency-exchange operators \cite{vasek2015there}, \emph{client-side hosted wallets}~\cite{mycelium-wallet,CarbonWallet,CitoWiseWallet,coinomi-wallet,InfinitoWallet} (and their sub-category of \emph{embedded wallets}~\cite{thirdweb-embedded-2024,beam-2024}) have started to proliferate.
In such wallets, the main functionality, including the storage of private keys, has moved to the user side;
hence, trust in the provider is reduced, but the users still depend on the provider's infrastructure.

To increase the security of former wallet categories, multi-factor authentication (MFA) is often used, which enables spending crypto-tokens only when a number of secrets are used together.
Wallets from a split control category \cite{eskandari2018first} provide MFA against the blockchain.
This can be achieved by \emph{threshold cryptography wallets} \cite{goldfeder2015securing,mycelium-entropy,zengo-2024}, \emph{multi-signature wallets} \cite{Armory-SW-Wallet,Electrum-SW-Wallet,TrustedCoin-cosign,copay-wallet}, and \emph{state-aware smart-contract wallets} \cite{TrezorMultisig2of3,parity-wallet,ConsenSys-gnosis}.
Nevertheless, these schemes impose additional usability implications, performance overhead, or cost of wallet devices.

In sum, while there exist studies categorizing cryptocurrency wallets with regards to security~\cite{2015-Bitcoin-SOK,eskandari2018first,2023-Huoy-Sok-wallets-security,erinle2023sok,2020-Karantias-sok-wallets}, there has not been any study that would deal with the classification of wallets based on locally-verified and blockchain-verified authentication factors and their interconnection, which is the motivation for our work.

\subsection{Contributions}
With the existing key management approaches in mind, we aim to distinguish the type of wallet authentication principle(s) and procedure with regard to the wallet factors and their centralized and/or decentralized verification.

\begin{enumerate}
	\item In particular, we propose a classification scheme for cryptocurrency wallets based on the authentication factors used for centralized and decentralized authentication.
	\item To define the classification of authentication schemes of wallets, we introduce two new notions:  $k$-factor authentication against the blockchain and $k$-factor authentication against the authentication factors.
	\item We further extend our classification to accommodate threshold-signature approaches and centralized services that sign or co-sign transactions.
	\item We extend the categorization of wallets from the previous works~\cite{eskandari2018first,2015-Bitcoin-SOK}, while we also apply our proposed classification to reviewed wallets and their types.
\end{enumerate}
 
	\section{Classification of Authentication Schemes}\label{sec:wallets-classification}
In this section, we introduce our classification scheme for cryptocurrency wallets.
We denote the user by $\mathbb{U}$.

\subsection{Classification}
We introduce the notion of $k$-factor authentication against the blockchain and $k$-factor authentication against the authentication factors.
Using these notions, we propose a classification of authentication schemes, and we apply it to examples of existing key management solutions (see \autoref{sec:soa-wallet-types} and \autoref{tab:wallets-state-of-the-art}).

In the context of the blockchain, we distinguish between k-factor authentication  \textit{against the blockchain} and k-factor authentication \textit{against the authentication factors} themselves from $\mathbb{U}$'s point-of-view.
For example, an authentication method may require $\mathbb{U}$ to perform 2-of-2 multi-signature in order to execute a transfer, while $\mathbb{U}$ may keep each private key stored in a dedicated device -- each requiring a different password.
In this case, 2FA is performed against the blockchain since all blockchain miners verify both signatures.
Additionally, one-factor authentication is performed once in each device of $\mathbb{U}$ by entering a password in each of them.
For clarity, we classify authentication schemes by the following:
\begin{equation}\label{eqn:auth-simple}
	\Bigg (
	Z
	+  X_1
	\big/ \ldots \big/
	X_Z
	\Bigg),
\end{equation}
where $Z \in \{0, 1, \ldots \}$, the first operand of ``+'', represents the number of authentication factors against the block\-chain and
$X_i \in \{0, 1, \ldots \} ~|~i \in [1,\ldots, Z]$, the second operand of ``+'',  represents the number of authentication factors against the i-th factor of $Z$ (i.e., local authentication to access a particular factor).
Hence, the ``+'' operator represents the connector between blockchain-verified factors and locally-verified factors.
With this in mind, we remark that the previous example provides $\left( 2 + 1/1 \right)$-factor authentication: twice against the blockchain (i.e., two signatures), once for accessing the first device (i.e., the first password), and once for accessing the second device (i.e., the second password).

\medskip
\subsubsection{\textbf{Extension for Threshold-Signature Co-Signing}}
Since the previous notation is insufficient for authentication schemes that use secret sharing \cite{shamir1979share}, we
extend it as follows:
\begin{equation}\label{eq:class-threshold}
	\Bigg (
	Z^{(W_1, \dots, W_Z)}
	+ \left( X_1^{1}, \ldots , X_1^{W_1} \right)
	\big/ \ldots \big/
	\left(X_Z^{1}, \ldots, X_Z^{W_Z} \right)
	\Bigg),
\end{equation}
where $Z$ has the same meaning as in the previous case,
$W_i \in \{0, 1, \ldots \}$ $|~i \in [1,\ldots, Z]$ denotes the minimum number of secret shares required to use the complete i-th secret $X_i$.
With this in mind, we remark that the aforementioned example provides $\left( 2^{(1, 1)} + (1)/(1) \right)$-factor authentication: twice against the blockchain (i.e., two signatures), once for accessing the first device (i.e., the first password), and once for accessing the second device (i.e., the second password).
We consider an implicit value of $W_i = 1$; hence, the classification $(2 + 1/1)$ represents the same as the previous one (the first notation suffices).
If one of the private keys were additionally split into two shares, each encrypted by a password, then such an approach would provide $\left( 2^{(2, 1)} + (1, 1)/(1) \right)$-factor authentication.

\medskip
\subsubsection{\textbf{Extension for Factors Provisioned by Centralized Service(s)}}
Since the previous scheme is not sufficient to express whether some factor verified at the blockchain was signed/co-signed/produced by a centralized service such as exchange or other (e.g., upon some off-chain authentication of $\mathbb{U}$), we extend the previous notation as follows:
\begin{eqnarray}\label{eq:class-threshold-centralized}
	&\Bigg (
	(Z-Y)^{(W_1, \dots, W_{Z-Y})}  \\
	+& \left( X_1^{1}, \ldots , X_1^{W_1} \right)
	\big/ \ldots \big/
	\left(X_{Z-Y}^{1}, \ldots, X_{Z-Y}^{W_{Z-Y}} \right) \\
	+& 	V_1	\big/ \ldots \big/ 	V_Y
	\Bigg),
\end{eqnarray}
where the first two lines have the same meaning as  \autoref{eq:class-threshold} with the only difference that $Z$ is ``decreased'' by $Y$ -- the number of blockchain-verified factors that were produced by a centralized party or more such parties (i.e., 3rd party wallet providers).
Therefore, the ``-'' operator puts blockchain-verified factors produced by $\mathbb{U}$ versus a centralized party(-ies) into a relation.
Thus, the 3rd line of \autoref{eq:class-threshold-centralized} expresses the number of factors $V_i$ that $\mathbb{U}$ have to present to a centralized party $i$ for successful authentication, ``authorizing'' it to use a blockchain-verified secret for $\mathbb{U}$-requested operation (e.g., a signature on $\mathbb{U}$'s transaction).

For example, if $\mathbb{U}$ logins to a centralized exchange by login/password and provides OTP for the execution of an external transaction, while the centralized exchange owns the private key used for signing a transaction, then such a scheme would provide $((1-1) + 2)$-factor authentication.
In another example, if $\mathbb{U}$ owns one private key in his local wallet (protected by a password) and requires a centralized party to make a multi-signature on her transaction (upon authentication by login/password + OTP), such a scheme would provide $((2-1) + 1 + 2)$-factor authentication.

\vspace{0.5cm}
\section{Review of Wallet Types}\label{sec:soa-wallet-types}
We extend the previous work of Eskandari et al. \cite{eskandari2018first} and Bonneau et al. \cite{2015-Bitcoin-SOK}, by categorizing and reviewing a few examples of key management solutions, while demonstrating the application of our classification (see \autoref{sec:wallets-classification}) to each wallet.
We remark that the categories are not necessarily disjoint and one wallet may thus belong to more than one category.

\subsection{Keys in Local Storage}
In this category of wallets, the private keys are stored in plaintext form on the local storage of a machine, thus providing $(1+0)$-factor authentication.
Examples that have historically enabled the use of unencrypted private key files are Bitcoin Core (until version 0.3) \cite{BitcoinCore}\footnote{Note that since Bitcoin Core enables also password-protected private keys, it also belongs to the next category.}, Electrum (before version 1.9)~\cite{Electrum-SW-Wallet} and MyEtherWallet (until 2018) \cite{MyEtherWallet}\footnote{Note that MyEtherWallet also enables to use password-protection of private keys, and thus it also belongs to the category of password-protected wallets. At the same time, this wallet belongs to the category of client-side hosted wallets since, besides browser extension (or locally ran DAPP), it can run from the server (which is the most common option).} wallets.
However, MyEtherWallet discouraged $\mathbb{U}$s from the local storage of private keys within the browsers since 2018 and mainly focused on integration with hardware wallets, while providing only a user interface for interaction with the hardware wallets.
Unencrypted private keys in Bitcoin Core (which comes only as a standalone application) were possible until version 0.3, however, later the wallet required password protection and enabled integration with hardware wallets through an external bridge called Hardware Wallet Interface~\cite{HWI-BTC-2024}.
Electrum wallet (which comes only as a standalone application) has followed Bitcoin Core and also enabled integration with hardware wallet and two-factor authentication.

\subsection{Password-Protected Wallets}
These wallets require $\mathbb{U}$-spe\-ci\-fied password to encrypt a private key stored on the local storage, thus providing $(1+1)$-factor authentication.
Examples that support this functionality are Armory Secure Wallet \cite{Armory-SW-Wallet}, Electrum Wallet \cite{Electrum-SW-Wallet}, MyEtherWallet \cite{MyEtherWallet}, Bitcoin Core \cite{BitcoinCore}, and Bitcoin Wallet \cite{BitcoinWallet}.
This category addresses physical theft, yet enables the brute force of passwords and digital theft (e.g., keylogger).

\subsection{Password-Derived Seed-Derived Wallets}
Password-derived and seed-derived wallets  (a.k.a., brain wallets and hierarchical deterministic wallets~\cite{maxwell2011deterministic,2012-hd-wallets}) can deterministically compute a sequence of private keys from a single password and or high-entropy seed, respectively.
This approach takes advantage of the key creation in the ECDSA signature scheme that many blockchain platforms use.
Examples of early password-derived wallets (for one of the configuration options) are Electrum \cite{Electrum-SW-Wallet}, Armory Secure Wallet \cite{Armory-SW-Wallet}, Metamask \cite{MetamaskWallet}, and Daedalus Wallet \cite{daedalus-wallet}.\footnote{Note that the password-based key derivation has been possible for one of the options or some versions.}
The wallets in this category provide $(1+X_1)$-factor authentication (usually $X_1 = 1$).
While hierarchical deterministic wallets with high enough seed entropy provide enough resistance to brute-forcing, password-derived wallets might suffer from weak passwords~\cite{courtois2016speed,vasek2017-brainwallets}.
Vasek et al.~\cite{vasek2017-brainwallets} found that most of the brain wallets with weak passwords in Bitcoin were drained within 24 hours from creation, but more likely in a few minutes.

\subsection{Hardware Storage Wallets}
In general, wallets of this category include devices that can only sign transactions with private keys stored inside sealed storage, while the keys never leave the device.
To sign a transaction, $\mathbb{U}$ connects the device to a machine and enters his passphrase.
When signing a transaction, the device displays the transaction's data to $\mathbb{U}$, who may verify the details.
Thus, wallets of this category usually provide $(1+1)$-factor authentication.
Popular USB (or Bluetooth) hardware wallets containing displays are offered by Trezor \cite{trezor-hw-wallet},  Le\-dger \cite{ledger-nano-s}, KeepKey \cite{keep-key}, and BitLox \cite{BitLox}.
An example of a USB wallet that is not resistant against tampering with $\mathbb{C}$ (e.g., keyloggers) is Ledger Nano \cite{ledger-nano}
-- it does not have a display, hence $\mathbb{U}$ cannot verify the details of transactions being signed.
An air-gapped transfer of transactions using QR codes is provided by ELLIPAL wallet \cite{ellipal-hw-wallet}.
In ELLIPAL, both $\mathbb{C}$ (e.g., smartphone App) and the hardware wallet must be equipped with cameras and display.
$(1+0)$-factor authentication is provided by a credit-card-shaped hardware wallet from CoolBitX \cite{CoolWalletS}.
A hybrid approach that relies on a server providing a relay for 2FA is offered by BitBox \cite{BitBox}.
Although a BitBox device does not have a display, after connecting to a machine, it communicates with $\mathbb{C}$ running on the machine, and at the same time, it communicates with a smartphone App through BitBox's server;
each requested transaction is displayed and confirmed by $\mathbb{U}$ on the smartphone.
One limitation of this solution is the lack of self-sovereignty.

\subsection{Split Control -- Threshold Cryptography}
In threshold cryptography \cite{shamir1979share,threshold-mackenzie2001two,threshold-gennaro2007secure,blakley1979safeguarding}, a key is split into several parties which enables the spending of crypto-tokens only when n-of-m parties collaborate.
Threshold cryptography wallets provide $\left( 1^{(W_1, \ldots, W_n)}\- + (X_1, \dots, X_n) \right)$-factor authentication, as only a single signature verification is made on a blockchain, but $n$ verifications are made by parties that compute a signature.
Therefore, all the computations for co-signing a transaction are performed off-chain, which provides anonymity of access control policies (i.e., a transaction has a single signature) in contrast to the multi-signature scheme that is publicly visible on the blockchain.
An example of this category is presented by Goldfeder et al. \cite{goldfeder2015securing}.
One limitation of this solution is a computational overhead directly proportional to the number of involved parties $m$ (e.g., for $m = 2$ it takes $13.26$s).
Another example of this category is a USB dongle called Mycelium Entropy \cite{mycelium-entropy}, which, when connected to a printer, generates triplets of paper wallets using 2-of-3 Shamir's secret sharing; providing $(1^{(2)} + (0, ~0))$-factor authentication.
A hybrid example from this category (and client-side hosted wallets) is Zengo Wallet~\cite{zengo-2024}, which uses 2-of-2 co-signing, where the $\mathbb{U}$ owns one key (protected by PIN) and Zengo server owns another key (protected by email/password and 3D face lock).
At the same time, $\mathbb{U}$ generates an encryption key that is used to backup the co-signing key at Zengo server, while the encryption key is stored at $\mathbb{U}$'s cloud provider, enabling $\mathbb{U}$ to recover the co-signing key. This example provides thus $(1^{(2)} + (1, ~2))$-factor authentication.

\subsection{Split Control -- Multi-Signature Wallets.}
In the case of multi-signature wallets, n-of-m owners of the wallet must co-sign the transaction made from the multi-owned address.
Thus, the wallets of this category provide $(n + X_1/\ldots/X_n)$-factor authentication.
One example of a multi-owned address approach is Bitcoin's Pay to Script Hash (P2SH).\footnote{We refer to the term  \textit{multi-owned address of P2SH} for clarity, although it can be viewed as Turing-incomplete smart contract.}
Examples supporting multi-owned addresses are Lockboxes of Armory Secure Wallet \cite{Armory-SW-Wallet} and Electrum Wallet~\cite{Electrum-SW-Wallet}.
A property of a multi-owned address is that each transaction with such an address requires off-chain communication.
A hybrid instance of this category and client-side hosted wallets category is Trusted Coin's cosigning service~\cite{TrustedCoin-cosign}, which provides a 2-of-3 multi-signature scheme -- $\mathbb{U}$ owns a primary and a backup key, while TrustedCoin owns the third key.
Each transaction is signed first by $\mathbb{U}$'s primary key and then, based on the correctness of the OTP from Google Authenticator, by TrustedCoin's key.
Therefore, this approach provides $(2-1) + 1 + 1$-factor authentication.
Another hybrid instance of this category and client-side hosted wallets is Bitpay Wallet (former Copay) \cite{copay-wallet}.
With Bitpay, $\mathbb{U}$ can create a multi-owned Bitpay wallet (for Bitcoin), where $\mathbb{U}$ has all keys in his machines and n-of-m keys co-sign each transaction.
Transactions are resent across $\mathbb{U}$'s machines during multi-signing through Bitpay.

\subsection{Split-Control -- State-Aware Smart Contracts}\label{sec:state-aware=contracts}
State-aware smart contracts provide ``rules'' for how owners can spend crypto-tokens of a contract, while they keep the current setting of the rules on the blockchain.
The most common example of state-aware smart contracts is the 2-of-3 multi-signature scheme that provides $(2+X_1/X_2)$-factor authentication.
An example of the 2-of-3 multi-signature approach that only supports Trezor hardware wallets is \textit{TrezorMultisig2of3} from Unchained Capital \cite{TrezorMultisig2of3}.
One disadvantage of this solution is that $\mathbb{U}$ has to own three Trezor devices, which may be an expensive solution that, moreover, relies only on a single vendor.
Another example of this category, but using the n-of-m multi-signature scheme, is Parity Wallet \cite{parity-wallet}.
However, two critical bugs \cite{parity-bug-July-17,parity-bug-November-17} have caused the multi-signature scheme to be currently disabled.
The n-of-m multi-signature scheme is also used in \textit{Gnosis Wallet} from ConsenSys \cite{ConsenSys-gnosis}.
The Gnosis multi-sig smart contract is also utilized in Bitpay wallet~\cite{copay-wallet} (for Ethereum).
A hybrid example of this category is Argent wallet~\cite{argent-wallet}, which runs as a smart contract that optionally enables to switch into 2-of-2 multi-signature mode, where one key is held in $\mathbb{U}$'s browser or smartphone App (protected by a password) and another key is stored at Argent's server, which co-signs $\mathbb{U}$ transaction upon successful email-based OTP verification, $((2-1) + 1 + 1)$-factor authentication.
In the case of a forgotten password, the Argent contract enables $\mathbb{U}$ (with only one private key) to switch off this feature after 7 days of inactivity.
Another example of this category are SmartOTPs~\cite{homoliak2020smartotps} that require the blockchain to verify signature in the first stage and One Time Password (OTP) in the second stage.
The signature is provided by a hardware wallet (protected by PIN), and OTP is provided by the authenticator device or the smartphone App (protected by a password/fingerprint).
Therefore, this solution also provides 2 + (1/1) authentication.
Moreover, SmartOTPs enable recovery of lost secrets by last resort address that can receive all funds of the wallets without any authentication upon elapsing a certain last resort timeout of inactivity (e.g., in months).

\setlength{\tabcolsep}{2.0pt}
\begin{table*}[!h]
	\scriptsize{
		\vspace{-0.6cm}
		\scalebox{0.81}{
			\begin{tabular}{lccccccccccccc}
				\toprule
				&
				\multicolumn{2}{c}{\specialcell{\textbf{Authentication Scheme}\\\\}} & \rotatebox[origin=l]{90}{\textbf{\specialcell{Air-Gapped Property}}} &
				\rotatebox[origin=l]{90}{\textbf{\specialcell{Resilience to\\Tampering w. Client}}} &
				\rotatebox[origin=l]{90}{\textbf{\specialcell{Post-Quantum\\Resilience}}} &
				\rotatebox[origin=l]{90}{\textbf{\specialcell{No Off-Chain\\Communication}}} &
				\rotatebox[origin=l]{90}{\textbf{\specialcell{Malware Resistance}}} &
				\rotatebox[origin=l]{90}{\textbf{\specialcell{Secrets Kept Offline}}} &
				\rotatebox[origin=l]{90}{\textbf{\specialcell{Independence of\\Trusted Third Party}}} &
				\rotatebox[origin=l]{90}{\textbf{\specialcell{Resilience to\\Physical Theft}}} &
				\rotatebox[origin=l]{90}{\textbf{\specialcell{Resilience to\\Loss of Secrets}}} &
				\rotatebox[origin=l]{0}{\textbf{\specialcell{Comments}}} \\
				\expandafter\cline\expandafter{\expandafter2\string-3\smallskip}
				&
				\textbf{Classification} &
				\textbf{Details} &
				& & & & & & & & &  \\
				\toprule
				
				\smallskip
				\textbf{Keys in Local Storage}                       & $1 + (0)$          & Private key                                                   &                                               &                                                         &                                                   &                                                   &                                             &                                                  &                                                        &                                                       &                                                      &                                                                                                                  \\
				Bitcoin Core \cite{BitcoinCore}                                & $1 + (0)$          & For one of the options                                                                      & N                                             & N                                                       & N                                                 & Y                                                 & N                                           & N                                                & Y                                                      & N                                                     & N/A                                                    &                                                                               \\
				Electrum Wallet \cite{Electrum-SW-Wallet}                             & $1 + (0)$          &   For one of the options                                                                    & N                                             & N                                                       & N                                                 & Y                                                 & N                                           & N                                                & Y                                                      & Y                                                     & N/A                                                    &                                                                               \\
				MyEtherWallet \cite{MyEtherWallet}                               & $1 + (0)$          &  For one of the options                                   & N                                             & N                                                       & N                                                 & Y                                                 & N                                           & N                                                & Y                                                      & N                                                     & N/A                                                    & \\
				\midrule
				\smallskip
				\textbf{Password-Protected Wallets}                  & $1 + (1)$          & \specialcell{Private key + encryption}                                  &                                               &                                                         &                                                   &                                                   &                                             &                                                  &                                                        &                                                       &                                                      & \\
				Armory Secure Wallet \cite{Armory-SW-Wallet}                         & $1 + (1)$          &                                                                       & N                                             & N                                                       & N                                                 & Y                                                 & N                                           & N                                                & Y                                                      & Y                                                     & N                                                    &                                                                               \\
				Electrum Wallet \cite{Electrum-SW-Wallet}                             & $1 + (1)$          &                                                                       & N                                             & N                                                       & N                                                 & Y                                                 & N                                           & N                                                & Y                                                      & Y                                                     & N                                                    &                                                                               \\
				MyEtherWallet \cite{MyEtherWallet}                     & $1 + (1)$          &                                                                       & N                                             & N                                                       & N                                                 & Y                                                 & N                                           & N                                                & Y                                                      & Y                                                     & N                                                    & \\
				Bitcoin Core \cite{BitcoinCore}                                 & $1 + (1)$          &                                                                       & N                                             & N                                                       & N                                                 & Y                                                 & N                                           & N                                                & Y                                                      & Y                                                     & N                                                    & \\
				Bitcoin Wallet \cite{BitcoinWallet}                                & $1 + (1)$          &                                                                       & N                                             & N                                                       & N                                                 & Y                                                 & N                                           & N                                                & Y                                                      & Y                                                     & N                                                    &\\
				\midrule
				\smallskip
				\textbf{Password-(/Seed-)Derived Wallets}                    & $1 + (X_1)$         &                                                                       &                                               &                                                         &                                                   &                                                   &                                             &                                                  &                                                        &                                                       &                                                                                           &                                                                               \\
				Armory Secure Wallet \cite{Armory-SW-Wallet}                         & $1 + (1)$          &                                                                       & N                                             & N                                                       & N                                                 & Y                                                 & N                                           & N                                                & Y                                                      & Y                                                     & Y                                                    & \\
				Electrum Wallet \cite{Electrum-SW-Wallet}                             & $1 + (1)$          &                                                                       & N                                             & N                                                       & N                                                 & Y                                                 & N                                           & N                                                & Y                                                      & Y                                                     & Y                                                    & \\
				Metamask \cite{MetamaskWallet}                             & $1 + (1)$          &                                                                       & N                                             & N                                                       & N                                                 & Y                                                 & N                                           & N                                                & Y                                                      & Y                                                     & Y                                                    & \\
				Daedalus Wallet \cite{daedalus-wallet}                             & $1 + (2)$          & \specialcell{2 passwords}                                          & N                                             & N                                                       & N                                                 & Y                                                 & N                                           & N                                                & Y                                                      & Y                                                     & Y                                                    & \\
				\midrule
				\smallskip
				\textbf{\specialcell{Hardware Storage Wallets}}             &        $1 + (X_1)$        &                                                                       &                                               &                                                         &                                                   &                                                   &                                             &                                                  &                                                        &                                                                                                            &                                      &                                                                               \\
				Trezor \cite{trezor-hw-wallet}                                      & $1 + (1)$          &                                                                       & N                                             & Y                                                       & N                                                 & Y                                                 & Y                                           & Y                                                & Y                                                      & Y                                                     & Y                                                    & \\
				Ledger \cite{ledger-nano-s}                                     & $1 + (1)$          &                                                                       & N                                             & Y                                                       & N                                                 & Y                                                 & Y                                           & Y                                                & Y                                                      & Y                                                     & Y                                                    & \\
				KeepKey \cite{keep-key}                                     & $1 + (1)$          &                                                                       & N                                             & Y                                                       & N                                                 & Y                                                 & Y                                           & Y                                                & Y                                                      & Y                                                     & Y                                                    & \\
				BitLox \cite{BitLox}                                       & $1 + (2)$          &  2 passwords$^*$                                                                     & N                                             & Y                                                       & N                                                 & Y                                                 & Y                                           & Y                                                & Y                                                      & Y                                                     & Y                                                    & \specialcell{~$^*$Additionally, protection\\~~against the evil maid attack}                                      \\
				CoolWallet S \cite{CoolWalletS}                       & $1 + (0)$          &                                                                       & N                                             & Y                                                       & N                                                 & Y                                                 & Y                                           & Y                                                & Y                                                      & P$^\dagger$                                                     & N/A                                                    & $^\dagger$\specialcell{Depending on the mode} \\
				
				Ledger Nano \cite{ledger-nano}                                 & $1 + (2)$          &         Password + GRID card                                                               & N                                             & N                                                       & N                                                 & Y                                                 & N                                           & Y                                                & Y                                                      & Y                                                     & Y                                                    & \\
				ELLIPAL wallet \cite{ellipal-hw-wallet}                                      & $1 + (1)$          &                                                                       & Y                                             & Y                                                       & N                                                 & Y                                                 & Y                                           & Y                                                & Y                                                      & Y                                                     & Y                                                    & \\
				BitBox USB Wallet \cite{BitBox}                          & $1 + (2)$          & 1 password and App                                                                       & N                                             & Y                                                       & N                                                 & Y                                                 & Y                                           & Y                                                & P$^\ddagger$                                                      & Y                                                     & Y                                                    & \specialcell{$^\ddagger$Requires a relay server}                \\
				
				\midrule
				\textbf{\specialcell{Split Control --\\Threshold Cryptography}}      & $1^{(W_1)} + (X_1^{1}, \ldots , X_1^{W_1})$         &                                                                       &                                               &                                                         &                                                   &                                                   &                                             &                                                  &                                                        &                                                       &                                                                                            &                                                                               \\
				Goldfeder et al. \cite{goldfeder2015securing}                            & $1^{(2)} + (1,1)$   & \specialcell{Assuming 2 devices, each\\protected by a password}               & N                                             & Y                                                       & N                                                 & N                                                 & Y                                           & N/A                                              & N/A                                                    & N/A                                                     & N/A &                                                                               \\
				Mycelium Entropy \cite{mycelium-entropy}                            & $1^{(2)} + (0,0)$   &                                                                       & N                                             & Y                                                       & N                                                 & N                                                 & Y                                           & Y                                                & Y                                                      & Y                                                     & N/A                                                    &                                                                               \\
				Zengo Wallet. \cite{zengo-2024}                            & $1^{(2)} + (1,2)$   & \specialcell{PIN at a device and password\\+ 3D face lock on the server}               & N                                             & N                                                       & N                                                 & N                                                 & N                                           & N                                              & N                                                    & Y                                                     & Y &  \specialcell{A hybrid client-side wallet.\\An encrypted backup on Zengo.\\Encryption key in cloud backup.}                                                                                \\  
				\midrule
				\smallskip
				\textbf{\specialcell{Split Control --\\Multi-Signature Wallets}}     & $Z + (X_1/\ldots/X_z)$    &                                                                       &                                               &                                                         &                                                   &                                                   &                                             &                                                  &                                                        &                                                                                                            &                                      &                                                                               \\
				\specialcell{Lockboxes of Armory\\Secure Wallet \cite{Armory-SW-Wallet}}             & $Z + (X_1/\ldots/X_z)$    & $Z$ up to 7, $X_i$ = 1                                                     & N                                             & Y                                                       & N                                                 & N                                                 & Y                                           & N                                                & Y                                                      & Y                                                     & N                                                    & \\
				Electrum Wallet \cite{Electrum-SW-Wallet}                             & $Z + (X_1/\ldots/X_z)$    & $Z$ up to 15, $X_i$ = 1                                                    & N                                             & Y                                                       & N                                                 & N                                                 & Y                                           & N                                                & Y                                                      & Y                                                     & Y                                                    & \\
				\specialcell{Trusted Coin's\\Cosigning Service \cite{TrustedCoin-cosign}}            &     $(2-1) + 1 + 1$           & \specialcell{2 private keys (one co-signed)\\ + 2 passwords and Google Auth.} & N                                             & Y                                                       & N                                                 & N                                                 & Y                                           & N                                                & N                                                      & Y                                                     & Y                                                    & A hybrid client-side wallet                                                         \\
				Bitpay Wallet (former Copay) \cite{copay-wallet}                                & $2 + (1/1)$        &  For one of the options (Bitcoin)                                                                     & N                                             & Y                                         & N                                                 & N                                                 & Y                                           & N                                                & P                                                      & Y                                                     & Y                                                   & A hybrid client-side wallet                                                          \\
				
				\midrule
				\smallskip
				\textbf{\specialcell{Split-Control --\\State-Aware Smart Contracts}} & $Z + (X_1/\ldots/X_z)$    &                                                                       &                                               &                                                         &                                                   &                                                   &                                             &                                                  &                                                        &                                                       &                                                      & \\
				\specialcell{TrezorMultisig2of3 \cite{TrezorMultisig2of3}}        & $2 + (1/1)$        & \specialcell{Assuming that each device\\is protected by a password}               & N                                             & Y                                                       & N                                                 & N                                                 & Y                                           & Y                                                & Y                                                      & Y                                                     & Y                                                                                &                                                                               \\
				Parity Wallet \cite{parity-wallet}                               & $Z + (X_1/\ldots/X_z)$    & $Z$ is unlimited, $X_i$ = 1                                                & N                                             & Y                                                       & N                                                 & Y                                                 & Y                                           & N                                                & Y                                                      & Y                                                     & Y                                                    & \\
				Gnosis Wallet \cite{ConsenSys-gnosis}                              & $Z + (X_1/\ldots/X_z)$    & $Z$ up to 50, $X_i$ = 1                                                    & N                                             & Y                                                       & N                                                 & Y                                                 & Y                                           & N                                                & Y                                                      & N/A                                                   & Y                                                    & \\
				Bitpay Wallet (former Copay) \cite{copay-wallet}                              & $Z + (X_1/\ldots/X_z)$    & \specialcell{For one of the options (Ethereum);\\$Z$ up to 50, $X_i$ = 1}                                                    & N                                             & Y                                                       & N                                                 & Y                                                 & Y                                           & N                                                & Y                                                      & N/A                                                   & Y                                                    & \specialcell{A hybrid client-side wallet.\\Uses Gnosis smart contract.} \\
				Argent Wallet\cite{argent-wallet}                              & $(2-1) + 1 + 1$    & \specialcell{For one of the options.\\Cosigning based on OTP.}                                                    & N                                             & N                                                       & N                                                 & N                                                 & Y                                           & N                                                & N                                                      & Y                                                   & Y                                                    & \specialcell{A hybrid client-side wallet.\\Uses Gnosis smart contract.} \\			
				SmartOTPs \cite{homoliak2020smartotps}                                         & $2 + (1/1)$        &  \specialcell{Private key and OTPs\\+ passwords}                                                                      & {Y}$^\circ$                                             & Y                                                       & {Y}                                                 & Y                                                 & Y                                           & Y                                                & Y                                                      & Y                                                     & 						Y    								& \specialcell{$^\circ$Fully air-gapped, if combined\\with ELLIPAL Wallet} \\
				
				\midrule
				\smallskip
				\textbf{Server-Side Wallets}                         &   $(1-1) + (X_1)$             &                                                                       &                                               &                                                         &                                                   &                                                   &                                             &                                                  &                                                        &                                                       &                                                      &                                      &                                                                               \\
				Coinbase \cite{CoinbaseWallet}, OKX~\cite{OKX-wallet}, Bitfinex~\cite{bitfinex-wallet}                                   & $(1-1) + (2)$          &    Password, Google Auth./SMS/passkeys                                                                    & N                                             & N                                                       & N                                                 & Y                                                 & N                                           & N                                                & N                                                      & Y                                                     & Y                                                    & \\
				Binance \cite{binance-exchange}                                   & $(1-1) + (3)$          &    Password, OTP from Google Auth./SMS/ + email                                                                    & N                                             & N                                                       & N                                                 & Y                                                 & N                                           & N                                                & N                                                      & Y                                                     & Y                                                    & \\
\midrule
				\textbf{Client-Side Wallets}                         &  $Z + (X_1)$          &                                                                       &                                               &                                                         &                                                   &                                                   &                                             &                                                  &                                                        &                                                       &                                                                                          &                                                                               \\
				BTC Wallet \cite{BTC-com-wallet}                                  & $1 + (2)$          &        Password + Biometrics or PIN                                 & N                                             & N                                                       & N                                                 & Y                                                 & N                                           & N                                                & N                                                      & Y                                                     & Y                                                    & Password-encrypted cloud backup.  \\
				Blockchain Wallet \cite{BlockchainInfoWallet}                           & $1 + (2)$          &   Password + Biometrics or PIN                                                                    & N                                             & N                                                       & N                                                 & Y                                                 & N                                           & N                                                & N                                                      & Y                                                     & Y                                                    & Password-encrypted cloud backup.  \\
				MyEtherWallet \cite{MyEtherWallet}                               & $1 + (1)$          &                                     & N                                             & N                                                       & N                                                 & Y                                                 & N                                           & N                                                & N                                                      & N                                                     & Y                                                    & \\
				Mycelium Wallet \cite{mycelium-wallet}                             & $1 + (1)$          &                                                                       & N                                             & N                                                       & N                                                 & Y                                                 & N                                           & N                                                & N                                                      & Y                                                     & Y                                                    &  \\
				CarbonWallet \cite{CarbonWallet}                                & $2 + (2)$          &   \specialcell{2 private keys stored in\\browser and smartphone}                                                                    & N                                             & Y                                                       & N                                                 & N                                                 & N                                           & N                                                & N                                                      & Y                                                     & Y                                                    & \\
				Citowise Wallet \cite{CitoWiseWallet}                             & $1 + (2)$          &                                                                       & N                                             & Y$^\P$                                                       & N                                                 & Y                                                 & N                                           & P$^\P$                                                & N                                                      & Y                                                     & Y                                                    &  \specialcell{$^\P$If combined with Trezor\\~or Ledger}               \\
				Coinomi Wallet \cite{coinomi-wallet}                              & $1 + (1)$          &                                                                       & N                                             & N                                                       & N                                                 & Y                                                 & N                                           & N                                                & N                                                      & Y                                                     & Y                                                    & \\
				Infinito Wallet \cite{InfinitoWallet}                              & $1 + (1)$          &                                                                       & N                                             & N                                                       & N                                                 & Y                                                 & N                                           & N                                                & N                                                      & Y                                                     & Y                                                    & 
				\\
				Harmony One \cite{harmony-one}                              & $1 + (2)$          &   \specialcell{For one of the options. OTP from Google Auth.\\ (verified at the blockchain) + password}                                                                    & Y                                             & N                                                       & N                                                 & Y                                                 & N                                           & N                                                & N                                                      & Y                                                     & Y                                                    & A hybrid smart contract wallet.
				\\
				\\
				\textbf{Embedded Wallets}                         &  $1 + (X_1)$          &                                                                       &                                               &                                                         &                                                   &                                                   &                                             &                                                  &                                                        &                                                       &                                                                                          &                                                                               \\
				Thirdwallet~\cite{thirdweb-embedded-2024}                              & $1 + (1)$          &     OAuth w. Google or OTP by email                                                                  & N                                             & N                                                       & N                                                 & Y                                                 & N                                           & N                                                & N                                                      & N                                                     & N                                                    & \specialcell{The key is stored in local storage\\of the browser in the plain text}
				\\
				Beam Wallet~\cite{beam-2024}                              & $1 + (2)$     &     \specialcell{Password and OAuth w. X or OTP by email}                                                                  & N                                             & N                                                       & N                                                 & Y                                                 & N                                           & N                                                & N                                                      & Y                                                     & N                                                    & \specialcell{The encrypted key is stored\\in local storage of the browser}
				\\
				
				\bottomrule
			\end{tabular}
		}
	}
	\caption{Comparison of state-of-the-art cryptocurrency wallets using our classification (see \autoref{sec:wallets-classification}) and other security features.}	
	\label{tab:wallets-state-of-the-art}
	\vspace{-0.3cm}
\end{table*}
\setlength{\tabcolsep}{1.4pt} 
\subsection{Hosted  Wallets}
Common features of hosted wallets are that they provide an online interface for interaction with the blockchain, managing crypto-tokens, and viewing transaction history. At the same time, they also store private keys on the server side.
If a hosted wallet has full control over private keys, it is referred to as a \textit{server-side wallet}.
A server-side wallet acts like a bank -- the trust is centralized.
Due to several cases of compromising such server-side wallets~\cite{binance-hack-2019,bitmart-hack-2021,ronin-hack-2022,coinex-hack-2023}, \cite{2018-coindesk-bithumb}, \cite{2014-Mt-Gox}, \cite{2016-Bitfinex-hack}, \cite{moore2013beware}, the hosted wallets that provide only an interface for interaction with the blockchain (or store only user-encrypted private keys) have started to proliferate.
In such wallets, the functionality, including the storage of private keys, has moved to $\mathbb{U}$'s browser (i.e., client).
We refer to these kinds of wallets as \textit{client-side wallets} (a.k.a., hybrid wallets \cite{eskandari2018first} and embedded wallets).

\myparagraph{\textbf{Server-Side Wallets}}
Coinbase \cite{CoinbaseWallet} is an early example of a server-side hosted wallet, which also provides exchange services.
Whenever $\mathbb{U}$ logs in or performs an operation, he authenticates himself against Coinbase's server using a password and obtains a code from Google Authenticator/Authy app/SMS/utilize passkeys.\footnote{When making an external transaction $\mathbb{U}$ does not have to provide OTP/passkeys and password again.}
Other examples of server-side wallets having similar security level to Coinbase are OKX~\cite{OKX-wallet} and Bitfinex~\cite{bitfinex-wallet}.
Another example is Binance~\cite{binance-exchange}, which, on top of the login/password, requires $\mathbb{U}$ to provide 2 OTPs to perform the external operation with the wallet -- one from the Google Authenticator and another one from the email on top of login and password.
The wallets in this category usually provide $((1-1) + 2)$-factor authentication when 2FA is enabled or $((1-1) + 3)$-factor authentication when 3FA is enabled.

\myparagraph{\textbf{Client-Side Wallets}}
An example of a client-side hosted wallet is Bitcoin (BTC) Wallet~\cite{BTC-com-wallet}, which requires (on top of encryption password) PIN or biometrics to access the wallet, and it provides 1-factor authentication against the blockchain.
Moreover, it enables cloud backup of private keys, encrypted by the user-specified (unrecoverable) password.
Equivalent functionality and security level as in BTC Wallet is offered by Blockchain DeFi Wallet \cite{BlockchainInfoWallet}, which
is an independent part of a combined client-side and server-side wallet.\footnote{Note that server-side wallet uses different key.} Other examples of this category are password-encrypted wallets, like Mycelium Wallet \cite{mycelium-wallet}, CarbonWallet \cite{CarbonWallet}, Citowise Wallet \cite{CitoWiseWallet}, Coinomi Wallet \cite{coinomi-wallet}, and Infinito Wallet \cite{InfinitoWallet}, which, in contrast to the previous examples, do not store backups of encrypted keys at the server.
A 2FA against the blockchain is provided in addition to password-based authentication, in the case of CarbonWallet.
In detail, the 2-of-2 multi-sig scheme uses the PC's browser and the smartphone's browser (or the app) to co-sign transactions.
The wallets in this category usually provide $(1+X_1)$-factor authentication, where $X_1$ is usually equal to 1 (in the cases of 2FA, it might be equal to 2).
Argent wallet~\cite{argent-wallet} is a wallet for Starknet (the L2 blockchain under Ethereum), which by default protects a private key stored in a browser by a password.
Nevertheless, it optionally provides even more flexible features thanks to smart contracts (see state-aware smart contract below).
Another example from this category is Harmony One wallet~\cite{harmony-one}, which focuses on usability of small amount transfers.
It requires $\mathbb{U}$ to login by a password, and then for each operation with the wallet she provides OTP from Google Authenticator only.
In contrast to many other solutions, OTP is verified at the blockchain as the only factor required to execute the operation.\footnote{Note that client stores a salt for OTP, increasing the security of OTP.}
Since $\mathbb{U}$ has no private key, the relaying services of Harmony make the signing of the transaction (to pay the fees).

\myparagraph{\textbf{Embedded Wallets}}
Embedded wallets can be viewed as a subcategory of client-side hosted wallets since they contain most of their features.
The only difference is that they do not need to run as a dedicated DAPP (on a new URL or locally as a browser extension) but they can run directly from the website of any service provider, and thus $\mathbb{U}$ does not need to leave its website nor interact with the browser extension.
An example of this category is Thirdweb embedded wallet~\cite{thirdweb-embedded-2024}, which enables $\mathbb{U}$ to login to the wallet (and create its local private key) either by 3rd party authentication method OAuth~\cite{OAuth} (through Google) or by sending OTP to $\mathbb{U}$'s email address.
In both cases, $\mathbb{U}$ needs to have access to the email address of her account, which usually requires at least the password to access the email (i.e., $(1+1)$-factor authentication).
However, since the key is stored at $\mathbb{U}$ device in plain text (in the case of browser within local storage), it is subject to extraction by malware, and moreover lacks the means for the recovery.
Another similar example is Beam wallet (with Join integrated)~\cite{beam-2024,beam-amazon-2024}, which in contrast to Thirdweb wallet, provides OAuth login through X or email-based OTP verification.
Nevertheless, it moreover enables encryption of locally stored private keys by a password (thus providing $(1+2)$-factor authentication).

\section{Security Features of Wallets}\label{sec:security}
\hfill

\noindent
We present a comparison of wallets and approaches from \autoref{sec:soa-wallet-types} in \autoref{tab:wallets-state-of-the-art}.
In detail, we apply our proposed classification on authentication schemes, while we also survey a few selected security properties of the wallets that also contain some features from the work of Eskandri et al. \cite{eskandari2018first}.
In the following, we briefly describe each property. 

\subsubsection{\textbf{Air-Gapped Property}}
We attribute this property (Y) to approaches that involve at least one hardware device storing secret information, which does not need a connection to a machine in order to operate.

\subsubsection{\textbf{Resilience to Tampering with the Client}}
We attribute this property (Y) to all hardware wallets that sign transactions within a device, while they require $\mathbb{U}$ to confirm transaction's details at the device (based on displayed information).
Then, we attribute this property to wallets containing multiple clients that collaborate in several steps to co-sign transactions (the chance that all of them are tampered with is low).

\subsubsection{\textbf{Post-Quantum Resilience}}
We attribute this property (Y) to approaches that utilize hash-based cryptography that is known to be resilient against quantum computing attacks~\cite{amy2016estimating}.

\subsubsection{\textbf{No Need for Off-Chain Communication}}
We attribute this property (Y) to approaches that do not require an off-chain communication/transfer of transactions among parties/devices
to build a final (co-)signed transaction, before submitting it to a blockchain (applicable only for $Z \geq 2$ or $W_i \ge 2$).

\subsubsection{\textbf{Malware Resistance (e.g., Key-Loggers)}}
We attribute this property (Y) to approaches that either enable signing transactions inside of a sealed device or split signing control over secrets across multiple devices.

\subsubsection{\textbf{Secret(s) Kept Offline}}
We attribute this property (Y) to approaches that keep secrets inside their sealed storage, while they expose only signing functionality.
Next, we attribute this property to paper wallets and fully air-gapped devices.

\subsubsection{\textbf{Independence of Trusted Third Party}}
We attribute this property (Y) to approaches that do not require a trusted party for operation, while we do not attribute this property to all client-side and server-side hosted wallets.
We partially (P) attribute this property to approaches requiring an external relay server for their operation.

\subsubsection{\textbf{Resilience to Physical Theft}}
We attribute this property (Y) to approaches that are protected by an encryption password or PIN.
We partially (P) attribute this property to approaches that do not provide password and PIN protection but have a specific feature to enforce the uniqueness of an environment in which they are used (e.g., Bluetooth pairing).

\subsubsection{\textbf{Resilience to Loss of Secrets}}
We attribute this property (Y) to approaches that provide means for the recovery of secrets (e.g., a seed of hierarchical deterministic wallets).

	\section{Conclusion}\label{sec:conclusion}
In sum, we proposed a classification of cryptocurrency wallets based on authentication methods and their factors.
In the classification, we distinguished between centralized (or local) authentication and decentralized authentication against the blockchain.
We demonstrated the application of our classification scheme in various categories and instances of the wallets that we moreover reviewed and cross-compared based on several security features from the literature. 
	\bibliographystyle{IEEEtran}
	\bibliography{ref}

\end{document}